\documentclass[a4paper,onecolumn,12pt,tightenlines,notitlepage,secnumarabic,showkeys,nofootinbib]{revtex4}

\usepackage{amssymb,amsmath,graphicx}

\usepackage{color}                       

\begin{document}

\title[CME Structure and Particle Acceleration]%
      {\Large CME Structure and Particle Acceleration}

\author{\bf Igor F. Nikulin}

\affiliation{
Sternberg Astronomical Institute (GAISh), Lomonosov Moscow State University,\\
Universitetskii prosp.\ 13, Moscow, 119992 Russia}

\email[E-mail: ]{ifn@sai.msu.ru}

\begin{abstract}
{\bf Abstract.}
The differences of the ejection-image structures in the chromospheric lines
and the coronal continuum are considered.
The outer, more diffusive scattering envelope of the ejection seems to be
produced by the excessive free electrons due to the Compton effect.
The resulting electric field accelerate the particles while the hard radiation
is acting.
The directed motion of the accelerated particles induces the magnetic fields.
\end{abstract}


\maketitle


\section{Introduction}\label{s:Intro} 

The coronal mass ejections (CME) are one of the most energetic phenomena
in the Solar system, which determine the space weather~\citep{Brueckner_74,%
Charbonneau_95}.
A cloud of the magnetized plasma with a mass of a few billion tones is ejected
from the coronal base into the interplanetary space with a velocity about
400--900~km/s due to some physical processes, which are still insufficiently
studied.
If the cloud is ejected in the direction to the Earth, then the magnetic
storms and aurorae can develop.
These phenomena can be associated both with the outbursts of
filaments/prominences and the flares. 
As usual, the speed of ejection in the case of flares are substantially
greater (up to 2000~km/s and even more).

\section{Structure and Dynamics of CME}\label{s:Struct}

%
\begin{figure} 
\centerline{\includegraphics[width=0.6\textwidth]{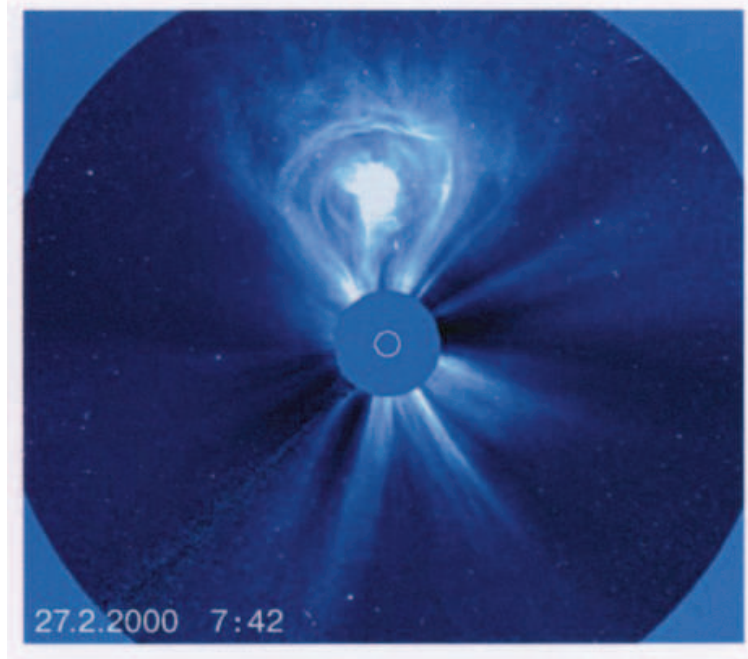}}
\caption{Outburst on 27 February 2000 with a well-developed envelope and
dense kernel in the middle (SOHO LASCO C3).}\label{fig:1}
\end{figure}

Apart from the insufficiently clear mechanism of acceleration of
the outbursts, there is some uncertainty in their characteristic structure.
This problem was unnoticeable when the outbursts of prominences and flares
were observed only from the Earth.
Detection of the outbursts by ground-based coronographs were very rare
and had a low quality, because of the considerable amount of scattered light.
However, launching the solar coronographs (OSO-7, SMM) beyond the
atmosphere~\citep{Howard_06} considerably increased the number of observable
outbursts, their contrast and quality, as well as have shown substantial
differences of the outburst structure in the chromospheric lines from
their pictures in the coronal continuum.
The coronal ejections of prominences and flares are typically associated with
the loop--arc structures, surrounding a kernel of the outburst at some distance
(Figure~\ref{fig:1}).
Such loops are seen in the pictures by spaceborne coronographs, such as
C2 and C3 SOHO, and the ground-based ones (for example, MLSO on 02.07.2015
and 15.01.2016); but they are not visible in the chromospheric lines,
\textit{e.g.}, Balmer lines of hydrogen or lines of neutral or ionized
helium.
These loops are sometimes diffuse and sometimes are composed of a few
separated layers, but most often they possess a weakly filamentary structure.

\begin{figure} 
\centerline{\includegraphics[width=0.6\textwidth]{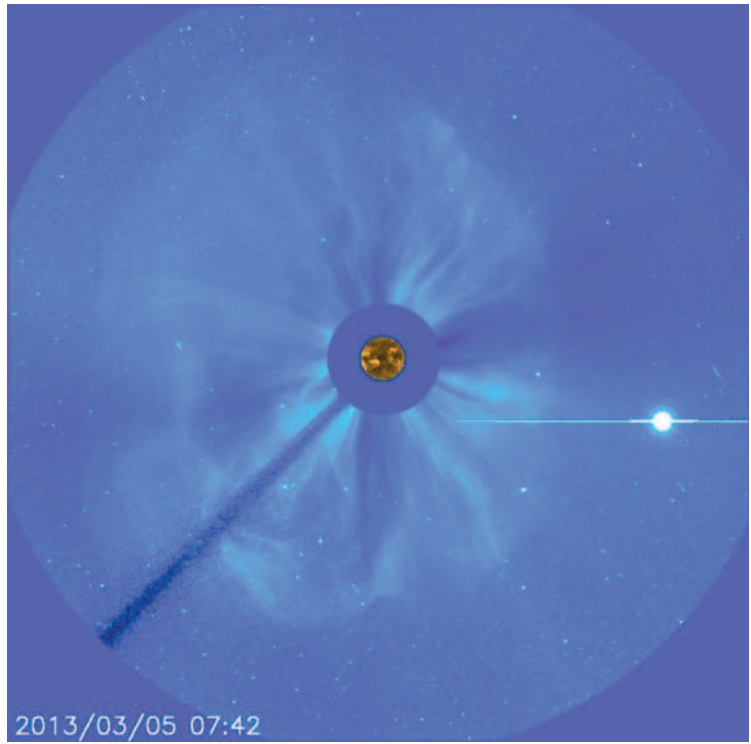}}
\caption{Envelope of the outburst on 05 March 2013, directed along
the Sun--Earth axis (SOHO L3).}\label{fig:2}
\end{figure}

Such loops are commonly interpreted as the magnetic ones, \textit{i.e.},
connecting the opposite magnetic regions in the region of outburst.
However, this is usually not confirmed by the detailed comparison with
magnetograms.
Moreover, the magnetic loops are not always located perpendicularly to
the line of sight and, consequently, should usually represent the elongated
eclipses (up to the stretching into a straight line).
Since this is usually not observed, one should assume that the outburst is
actually surrounded by a dynamic envelope, like a bubble, rather than by
the loops.
Such a conclusion is well confirmed by observations of the coronal outbursts
directed both to and away from the Earth (Figure~\ref{fig:2}), \textit{i.e.},
along the Sun--Earth axis (see also the images on 21.06.2015, 10.09.2014,
05.03.2013, and 04.11.2013).
So, one can see a concentric envelope (halo) about the Sun, which moves
away with the outburst and expands, but is always located at some distance
from it.

%

The above-mentioned general pattern is further complicated by the fact that
the outburst seen in the chromospheric lines often possesses a loop structure.
Then, it has the additional differences from the outer envelope, namely,
the outburst itself (kernel) is localized inside the envelope and lower,
there are clearly expressed structural features (but they do not have
the shape of a bubble), and its speed is a bit less.

How can such a loop be formed, and what processes are responsible for this?
Why is it located at a substantial distance from the outburst, and how is
this distance determined? Why is it not seen in the chromospheric lines?
Why are the structures of the outburst and its envelope so different?
The answers to these questions are important not only for the theory of
nonstationary solar processes and the formation of interplanetary medium.
It is surprising that the dramatic difference in the images of outbursts
in the lines and the coronal continuum was not paid attention before.
So, how can we explain this difference in the images, why is the outer
envelope of the outburst not seen in the chromospheric lines, and why is it
often more diffuse?

\section{The Mechanism}\label{s:Mech}

As follows from observations, the ejections are always associated with
a considerable burst of hard-ray emission (from keV to MeV and sometimes
even up to the gamma-radiation).
The especially large and hard fluxes of the pulsed radiation usually
follow the flares.
Leaving apart the problem of origin of this radiation, let us consider
its interaction with the environmental medium, which is a hydrogen--helium
plasma with small admixture of other elements.
The harder is the incident radiation, the stronger will be the Compton
effect, \textit{i.e.}, transfer of some part of the energy and momentum
of the incident quantum to free or weakly-bound plasma electrons.
The corresponding Compton electrons will be ejected forward (\textit{i.e.},
from the photosphere, in the same direction with incident radiation);
while protons, which are three orders of magnitude more massive, should remain
behind due to inertia.
As a result, the electric field will develop.
Such a process is inevitable under the presence of hard emission and
the specified direction of its propagation~\citep{Evans_72}.

%
\begin{figure} 
\centerline{\includegraphics[width=0.9\textwidth]{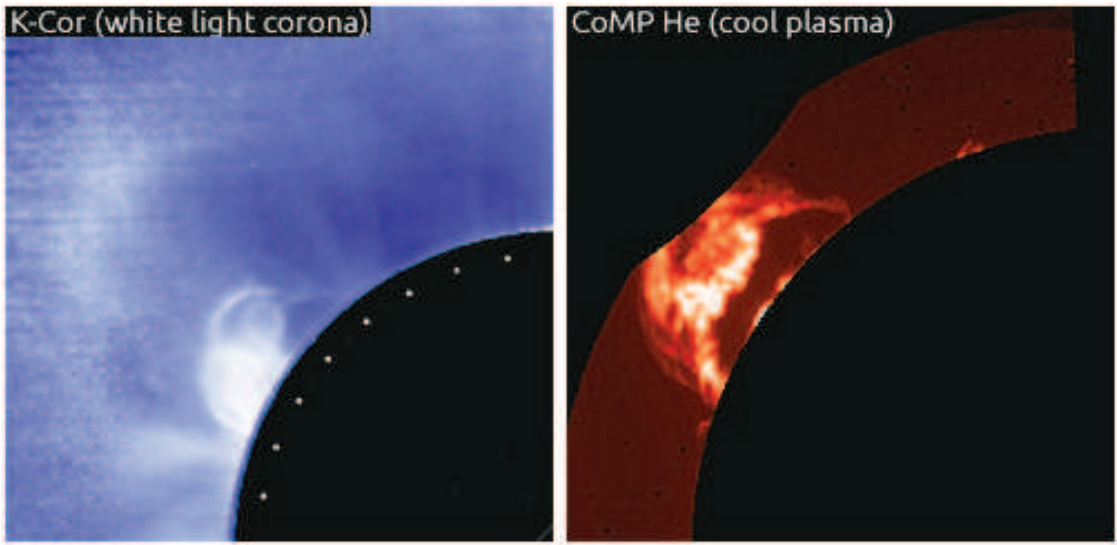}}
\caption{Outburst on 02 July 2015 in the coronal continuum (left panel)
and in the line of helium (right panel). Onset of formation of the envelope
can be noticed in the left panel.}\label{fig:3}
\end{figure}

Evidently, the larger is the flux and harder the radiation, the stronger will
be the field.
A gap with an electric field should appear: the outer region (envelope) should
be charged negatively; and the region near the photosphere (the outburst and
below), positively.
This field will not only accelerate the protons outward but also decelerate
the electrons and prevent them from the escape, despite of accelerating action
by the hard radiation.
An excess of the free electrons will be formed in the walls of the bubble.
As a result, its brightness should quickly increase, as is seen, for example,
in the film by the ground-based coronograph MLSO/HAO/KSOR of the event on
02.07.2015 (\url{https://www2.hao.ucar.edu/mlso/gallery/3-part-cme-2}).
A pair of frames from this film is presented in Figure~\ref{fig:3}.
One can see only a top part of the outburst in the beginning of the event.
Subsequently, a diffuse boundary appears at some distance in front of it.
This boundary moves away from the outburst, becomes more contrast and
brighter in the course of time.
The coronal magnetic fields---already existed before the outburst and
compressed by the kinetic energy of Compton electrons---can play an important
role in trapping the electrons and forming the bubble walls.
Effect of the hard radiation and accelerated electrons on the surrounding
structures is detectable by bending the neighbouring coronal rays outward from
the burst as well as by the sharper boundary of the rays from that side.
The accelerating electric field will exist while the hard radiation is
present.
Such acceleration mechanism is justified by perfect identity between
the temporal structures of the X-ray and microwave emission by
the flares~\citep{McLean_85}, \textit{i.e.}, the rate of electron acceleration
(characterized by the microwaves) is completely determined by the rate of
energy input (characterized by the X-rays).

The Compton effect is most expressed in the range of radiation
energy~0.01--5.0~MeV; the corresponding interval being wider for the light
elements~\citep{Evans_72}.
The photoelectric effect prevails at the less energies, and the creation of
pairs at the greater energies.
The above-mentioned energy range is quite typical for the X-ray solar flares
of classes~C--X.

%
\begin{figure} 
\centerline{\includegraphics[width=1.0\textwidth]{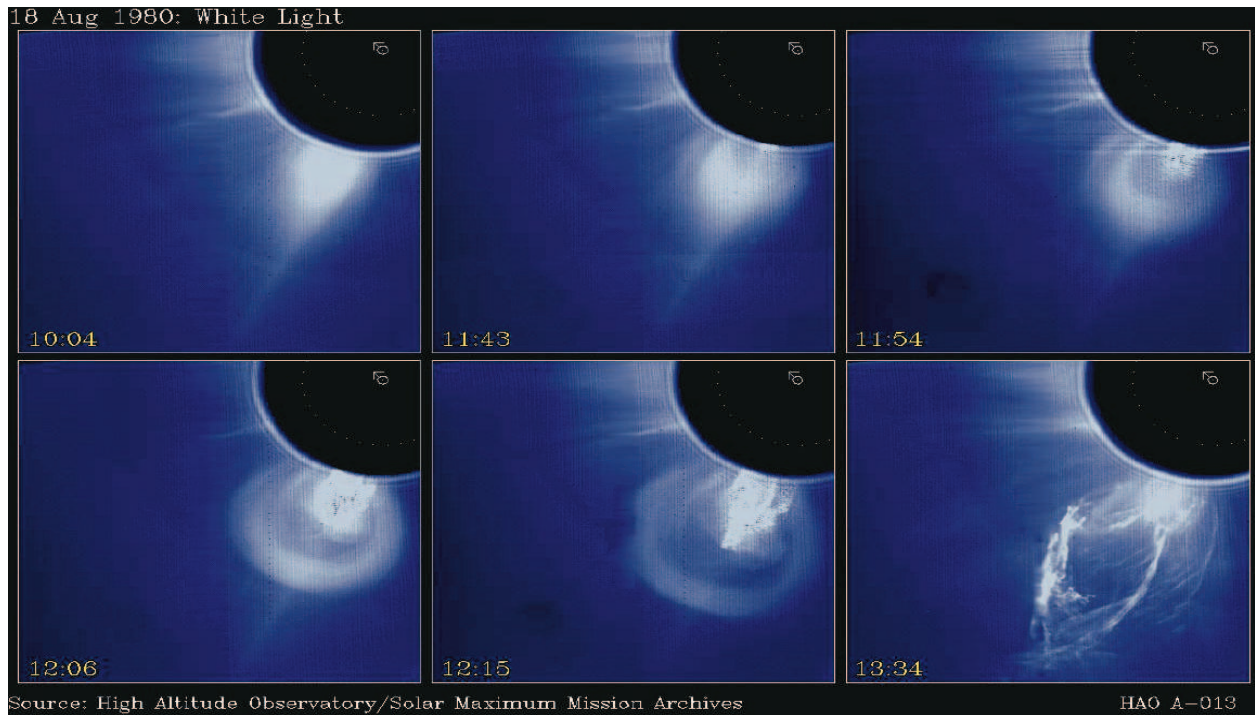}}
\caption{Development of the outburst on 18 August 1980 (SMM). The outer
diffuse envelope of the approximately circular shape is well seen.}
\label{fig:4}
\end{figure}
 
These processes are well illustrated by the frame sequence in one of
the first extra-atmospheric observations of CME by SMM apparatus
(Figure~\ref{fig:4}).
The ejection itself took place on 18.08.1980. 
An obscuring disc of the coronograph covered only 160{\%} of the solar
diameter, \textit{i.e.}, the outburst could be observed much closer to the Sun
than by LASCO SOHO without C1 (C2~-- $ 2R_{\odot} $,
C3~-- 3.7$ 2R_{\odot} $). 
In the first frame (10:04~UT), its visible part reminds a fan, and
the disturbances of the neighbouring coronal structures are already
noticeable.
In the second and third frames (11:43 and 11:54~UT), one can see formation
of the envelope and growth of its density, appearance of the top part of
the outburst, and formation of a gap between the outburst and the envelope.
In the fourth and fifth frames, the envelope and outburst are separated most
clearly; the structure of the outburst being much more contrast than
the envelope.
One more difference between the outburst and envelope is that the shape of
envelope is approximately circular, and its structure is uniformly diffuse.
(However, it should be kept in mind that the quality of SMM images was
less than in SOHO).
In the last frame, taken over an hour later, the envelope already dissipated,
and structure of the flying outburst is well seen.

The diverse structure and dynamics of CME envelopes still have to be studied
in more detail, since their powerful electrodynamic processes determine
not only shape of the bubble but also the velocity and direction of motion of
its details as well as the structure and intensity of magnetic fields in
the envelope and physical parameters of the shock wave at front of
the outburst.
As suggested by frontal parts of the envelopes (which are often very
nonuniform), the primary wave of electrons---usually corresponding to
the most powerful X-ray pulse---can break away from the envelope and go into
the interplanetary space, using the negatively-charged envelope as
an accelerating shield.

So, why is the envelope of the outburst not visible in the chromospheric
lines?
In the case of narrow-band line observations, Thomson radiation corresponding
to this band (\textit{i.e.}, the radiation by the outburst envelope scattered
by free electrons) should be much weaker, $ 10^{-5}-10^{-6} $ times
\citep{Allen_73}, and therefore it will be unnoticeable.
On the other hand, coronographs recording the coronal continuum integrate
the radiation scattered by electrons over a wide spectral range, up to
2000{\AA} \citep{Morrill_06}.
So, this accumulated radiation can exceed the narrow-band emission in
the strong chromospheric lines, resulting in the considerable difference
between the images of outbursts in the coronal continuum (C2, C3 SOHO) and
in the lines.
This becomes especially noticeable after injection of additional electrons
into the envelope by the X-rays.
Therefore, only the outburst itself will be seen in the emission of lines,
while one can see also the surrounding envelope in the pictures from
coronographs.

\section{Magnetic Fields}\label{s:Magnet}

Unfortunately, the problem of origin of the magnetic fields in space,
in general, and in the Sun, in particular, was studied by now only from
the viewpoint of amplification by various theoretical mechanisms of
the fields already existing and rising from the solar interiors rather than
a creation of the field \textit{in situ}, which is required by the short
characteristic times of variation of the observed magnetic fields.
Evidently, a presence of the electric currents---a directed motion of
the electric charges---is a necessary prerequisite for the formation of
magnetic fields.
The mechanism of creation of the electric fields with the acceleration times
corresponding to the characteristic scales of variation in the fluxes of
hard emission ($ 10^{-1}-10^{-3} $~s), as described above, assumes the minimal
times of acceleration of the particles, formation of the electric currents
and the respective magnetic fields.
This can show a way for explanation of many problems of variation of the solar
magnetic fields, especially, in the dynamic processes.

\section{Discussion}\label{s:Discus}

\begin{figure}[h]
\centerline{\includegraphics[width=0.3\textwidth]{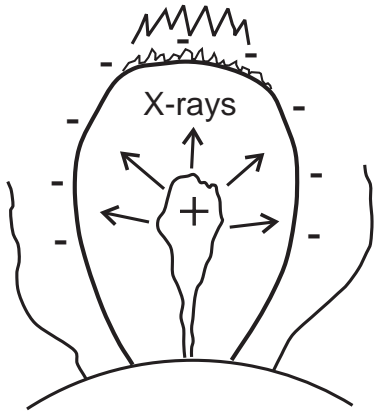}}
\caption{Sketch of formation of the electric field and the outer envelope
of the outburst, corresponding to Figure~\ref{fig:1}.}\label{fig:5}
\end{figure}

The observations of outbursts by coronographs enable us---by their comparison
with ground-based or space-born hydrogen and helium images---to identify
the regions of intensive Thomson scattering, \textit{i.e.}, where
the electrons prevail in the envelope.
Such observations can be performed only by the wide-band coronographs
(K-corona).
However, there are some problems with observation of the initial phase of
the outburst, before it goes beyond the edge of the coronograph's obscuring
disc.
This phase can be observed at the solar disc or near the limb in UV lines,
but there might be some difficulties in searching for a unique correspondence.
The unobservability of the bubbles (electron envelopes of the outbursts) in
the chromospheric and UV lines is a clear evidence for another origin of
their emission.
Therefore, the should be a favourable opportunity to separate the physically
different zones, namely, the region of Thomson scattering by free electrons
and plasma emission in the lines with the electric field between these zones
(Figure~\ref{fig:5}).

%

In fact, this is the first case when, by comparing the images of outbursts
in the coronal emission and filterheliograms or spectroheliograms,
it is possible to detect a presence of the extended electric fields in
the very remote object, such as the Sun.
As a result, one can determine a localization of the particle-acceleration
regions as well as  the mechanism of their acceleration, namely, by
the electric field.
The problem of particle acceleration is ultimately reduced to the question
how form a charge separation for a sufficiently long time?
In the case under consideration, such separation is supported by the energy
of hard radiation of the outburst.
This mechanism of particle acceleration seems to be suitable for all
the processes involving the directed fluxes of hard radiation.
Of course, the so clear pattern of separation of the envelope and
the outburst, as in Figures~\ref{fig:1} and~\ref{fig:4}, is not always
observed.
It depends on the power and temporal structure of the X-ray bursts,
surrounding coronal structures, and the magnetic fields, which can restrict
the solid angle of the outburst.
So, the above-mentioned separation is the result of equilibrium between
the pushing field of radiation and the electric field contracting this
gap.

Thereby, it becomes possible to explain the old puzzle discussed, for example,
in the book by \citet{Waldmeier_41}: Why do the outbursts with velocities much
less than parabolic leave the Sun?
This refers, first of all, to the ejection of prominences whose initial
velocities are usually small and increase quite slowly, as was demonstrated
in the above-cited book by the results of many researchers.
In that time, there were attempts to explain such a fact by the radiation
pressure.
So, one can say that this idea is reborn now at a new level.
The case is that, under the presence of electric field, the outbursts get
not a solitary pulse, resulting in the escape from the Sun, but rather
a continuous traction in this field, which gradually brings them
into the escape zone during the period of action of the hard radiation.
As distinct from the flares, the ejection velocities of the prominences
are usually much less, especially in the beginning of the process.
This seems to be explained by the relatively weak X-rays during ejection
of the prominence and, respectively, by the weak accelerating field.
In general, such cases are typical for the ejection of the prominences;
and they are described, for example, in our paper \citep{Nikulin_06} by
the data of original observations.

So, it seems rather plausible that the \citet{Moreton_60} waves, often
formed during the powerful solar flares, represent just a trace of
the expanding base of the electron envelope of outburst on the solar surface.
In some sense, this is in agreement with the hypothesis by \citet{Uchida_68}
that this wave represents the bottom part of the coronal MHD shock wave
propagating along the solar surface.

\section{Conclusions}\label{s:Concl}

\begin{enumerate}

\item
There is a considerable difference in the structure of CME observed in
the chromospheric lines and in the coronal continuum.
This difference seems to be caused by the escape of electrons due to
the Compton effect into the outer envelope, concentric with respect to
the outburst, and the respective amplification of Thomson scattering
in this region.

\item
To accelerate the charged particles, an initial energy in the form of
a hard-energy radiation pulse in the range 0.01--5.0~MeV is necessary.

\item
The corresponding directed emission can produce by the Compton effect
(the ejection of electrons and a delay of protons due to their inertia)
the accelerating electric field, whose polarity is positive near
the radiation source and negative in the envelope.

\item
The intensity and the range of influence of the electric field is
proportional to the intensity and energy of X-ray emission.

\item
Lifetime of the electric field should be immediately related to
the period of action of the hard radiation.

\item
The presented scheme of the particle acceleration seems to be applicable
not only for the processes in Sun but also for any space objects possessing
the fluxes of hard radiation.

\end{enumerate}

\section*{Acknowledgements}

I am grateful to the teams constructed and served the spacecraft whose
observational data, stored in the public repositories, were used in
the present paper.

This work was supported by the Russian Foundation for Basic Research,
grant no.~16-02-00585.

At last, I am grateful to Yu.V.~Dumin for his valuable help in the translation
and preparation of the manuscript.


\end{document}